\documentstyle[aps,prb,epsfig,twocolumn,floats]{revtex}
\begin{document}
\draft
\title{The effect of chain stiffness on the phase behaviour of isolated homopolymers}
\author{Jonathan P.~K.~Doye, Richard P.~Sear and Daan Frenkel}
\address{FOM Institute for Atomic and Molecular Physics, 
Kruislaan 407,\\ 1098 SJ Amsterdam, 
The Netherlands}
\date{\today}
\maketitle
\begin{abstract}
We have studied the thermodynamics of isolated homopolymer chains of 
varying stiffness using a lattice model. A complex phase behaviour is 
found; phases include chain-folded `crystalline' structures, 
the disordered globule and the coil.  It is found, in agreement with 
recent theoretical calculations, that the temperature at which the 
solid-globule transition occurs increases with chain stiffness, whilst 
the $\theta$-point has only a weak dependence on stiffness.  
Therefore, for sufficiently stiff chains there is no globular phase
and the polymer passes directly from the solid to the coil.
This effect is analogous to the disappearance of the liquid phase observed 
for simple atomic systems as the range of the potential is decreased.
\end{abstract}
% 61.25.Hq  Macromolecular and polymer solutions; polymer melts; swelling
% 36.20.-r  Macromolecules and polymer molecules
%          (for polymer reactions and polymerization, see 82.35; for
%          biological macromolecules and polymers, see 87.15)
% 07.05.Tp  Computer modeling and simulation
\section{Introduction}
Most of the work on the thermodynamics of isolated homopolymers has
concentrated on the collapse transition between the high temperature 
coil and the lower temperature dense globule. 
This emphasis is a result of the extensive theoretical work on the
transition\cite{Flory67,deGennes79} and the relative ease of simulations at these 
low densities---indeed simulations have been performed up to very large 
sizes,\cite{Grassberger95a} thus providing detailed tests of 
theoretical predictions. 

In contrast, there has been much less interest in the possibility of a 
low temperature order-disorder transition between two dense phases both 
because of the greater difficulty of simulating dense polymers and 
because fewer theoretical expectations for such a 
transition are available. 
This situation contrasts with that for heteropolymers where, in the context of
protein folding, there has been intense interest in the transition between 
the molten globule and the native state of the protein. 
The structure of the native state reflects the amino acid sequence and the 
specific interactions between these units, and so it might be thought that an
ordered structure is less likely when all polymer units are identical.
However, this is not the lesson from other finite systems. 
For example, homogeneous atomic clusters show a rich low temperature phase 
behaviour; there is the finite-size analogue of the first-order melting 
transition,\cite{Berry88a,Labastie90a} surface melting,\cite{Nielsen94a} and 
even low temperature transitions between different ordered forms.\cite{Doye97d} 

Recently, the existence of an isolated homopolymer order-disorder transition has 
begun to be confirmed.\cite{Bradley93a,Kuznetsov96a,Zhou96a}
In their study of a lattice homopolymer model which involved three-body forces
Kuznetsov {\it et al.\/} found phases with orientational 
order,\cite{Kuznetsov96a} and in their simple off-lattice model 
Zhou {\it et al.\/} observed a order-disorder
transition and also a solid-solid transition.\cite{Zhou96a} 
Moreover, in some earlier studies of the collapsed polymer glimpses of these
transitions were seen.\cite{Kolinski86a,Kolinski86b,Kolinski87a,Szleifer92a}
Here, we add to this growing understanding of the low temperature phase
behaviour of isolated homopolymers by studying a lattice model of a 
semi-flexible polymer, looking particularly at the effect of stiffness on the 
order-disorder transition. We compare our results with the phase diagram 
calculated in recent theoretical\cite{Doniach96a} and 
simulation\cite{Bastolla97a} studies of the polymer model we use here.

\section{Methods}
\subsection{Polymer Model}
\label{sect:pmodel}
In our model the polymer is represented by an $N$-unit self-avoiding walk 
on a simple cubic lattice. 
There is an attractive energy, $\epsilon$, between non-bonded polymer units on 
adjacent lattice sites and an energetic penalty, $\epsilon_g$, for kinks in the 
chain. The total energy is given by
\begin{equation}
E=-n_{pp}\epsilon+n_g\epsilon_{g}
\label{eq:energy}
\end{equation}
where $n_{pp}$ is the number of polymer-polymer contacts
and $n_g$ is the number of kinks or `gauche bonds' in the chain.
$\epsilon$ can be considered to be an effective interaction representing 
the combined effects of polymer-polymer, polymer-solvent and solvent-solvent 
interactions, and so our model is a simplified representation of a semi-flexible 
polymer in solution. 
The behaviour of the polymer is controlled by the ratio $kT/\epsilon$; large 
values can be considered as either high temperature or good solvent conditions, 
and low values as low temperature or bad solvent conditions.
The parameter $\epsilon_g$ defines the stiffness of the chain. 
The polymer chain is flexible at $\epsilon_g$=0 and becomes stiffer as 
$\epsilon_g$ increases.  In this study we only consider $\epsilon_g\geq 0$.

When $\epsilon_g=0$, this model corresponds to one originated by Orr\cite{Orr47a}
and has been much used to study homopolymer 
collapse.\cite{Grassberger95a,Szleifer92a,Bruns84a,Meirovitch90a,Tesi96a}
The system with positive $\epsilon_g$ has been recently studied 
theoretically by Doniach {\it et al.\/}\cite{Doniach96a} and using simulation by
Bastolla and Grassberger.\cite{Bastolla97a} 
In our work we pay special attention to the structural changes of a polymer of a 
specific size. In this sense our work is complementary to that of 
ref.~\onlinecite{Bastolla97a} which focussed on the accurate mapping of 
the phase diagram.

Our model was chosen because we wished to have the simplest model in
which we could understand the effects of stiffness, and because it gives 
low energy states that have chain folds resembling those found in
lamellar homopolymer crystals.\cite{Keller68a}  
Similar structures have previously been seen in a diamond-lattice model
of semi-flexible polymers,\cite{Kolinski86a,Kolinski86b,Kolinski87a} 
and in molecular dynamics simulations of isolated polyethylene 
chains.\cite{Kavassalis93a,Sundararajan95a,Fujiwara97a}
Particular mention should be made of the studies by Kolinski {\it et al.\/} on
the effect of chain stiffness on 
collapse\cite{Kolinski86a,Kolinski86b,Kolinski87a} since they found evidence 
for many of the phenomena that we explore systematically here. 

The global minimum at a particular (positive) $\epsilon_{g}$ is 
determined by a balance between maximizing $n_{pp}$
and minimizing $n_{g}$. 
If the polymer is able to form a structure that is a cuboid with
dimensions $a\times b \times c$ ($N=abc$), where
$a\le b \le c$, then 
\begin{equation}
n_{pp}=2N-ab-ac-bc+1
\end{equation}
and
\begin{equation}
n_{g}^{min}=2ab-2
\label{eq:ng}
\end{equation}
The structures that correspond to $n_g=n_g^{min}$ have the polymer chain
folded back and forth along the longest dimension of the cuboid.
By minimizing the resulting expression for the energy one
finds that the lowest energy polymer configuration should have 
\begin{equation}
a=b \quad\hbox{and}\quad {c\over a}=1+{2\epsilon_{g}\over\epsilon}.
\label{eq:aspect}
\end{equation}
Therefore, at $\epsilon_{g}=0$ the ideal shape is a cube and for 
positive $\epsilon_{g}$ a cuboid extended in one direction, 
the aspect ratio of which increases as the chain becomes stiffer.
As the ideal aspect ratio of the crystallite is independent of $N$, 
its squared radius of gyration, $R_g^2$, will scale as $N^{2/3}$; 
this scaling is the same as for the disordered collapsed globule.

\begin{figure}
\epsfig{figure=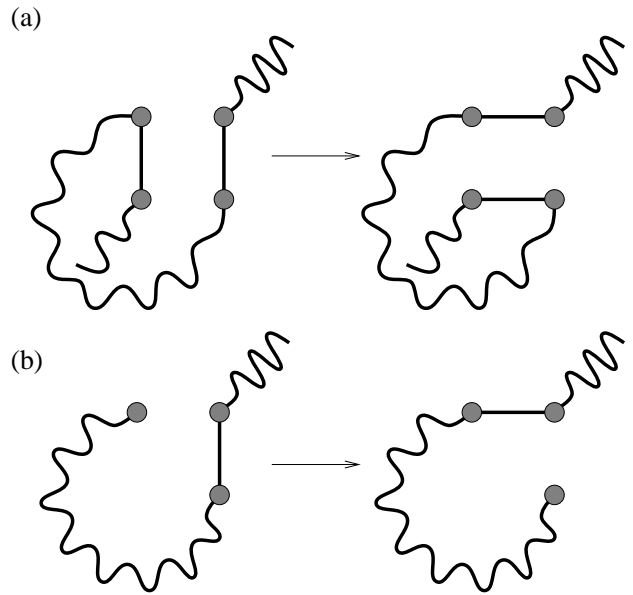,width=8.2cm}
\vspace{3mm}
\caption{Bond-flipping moves. (a) Four-bond flip in which the mid-section of the 
chain is reordered.  Moves of this type were only attempted if they preserved 
the integrity of the chain.
(b) Two-bond flip which results in a new chain end.
Only the bonds which change are explicitly depicted; 
other sections of the polymer are represented by wiggly lines.}
\label{flips}
\end{figure}

Substitution of the optimal dimensions of the cuboid [Equation (\ref{eq:aspect})] 
into Equations (\ref{eq:energy})--(\ref{eq:ng}) gives a lower bound to the 
energy of the global minimum,
\begin{equation}
E_{opt}/\epsilon=-2N+3N^{2/3}(1+2\epsilon_g/\epsilon)^{1/3}
-1-2\epsilon_g/\epsilon.
\label{eq:lbound}
\end{equation}
However, at most sizes and values of $\epsilon_{g}$ it is not possible 
to form a cuboid with the optimal dimensions, and so the energy of the 
global minimum will be higher than given by the above expression.
Nevertheless, it is easy to find the global minimum just 
by considering the structures which most closely approximate this ideal shape.

\subsection{Simulation Techniques}

Recent advances in simulation techniques have made it possible to begin to 
study dense polymer systems. In particular, we use configurational-bias
Monte Carlo\cite{Siepmann92a} including moves in which a mid-section 
of the chain is regrown.\cite{Dijkstra94a} 
We also make occasional bond-flipping moves [Figure \ref{flips}] which, 
although they do not change the shape of the volume occupied by the polymer, 
change the path of the polymer through that 
volume.\cite{Ramakrishnan97a,Deutsch97a}
These moves speed up equilibration in the dense phases.
The simulation method was tested by comparing to results obtained for 
$\epsilon_g=0$ by exact enumeration.\cite{Ishinabe94a,Douglas97a}
Thermodynamic properties, such as the heat capacity, were calculated 
from the energy distributions of each run using the 
multi-histogram method.\cite{Ferrenberg89a,Poteau94a}

The number of Monte Carlo steps used in our simulations 
was typically 4--30$\times 10^6$. 
The longer simulations were required for the larger polymers, 
especially at temperatures where two or more states coexisted. 
The sizes of the polymers we studied were $N$=27, 64, 100, 216, 343. 
These sizes are `magic numbers' for $\epsilon_g=0$, because 
compact cuboids of low aspect ratios can be formed at these sizes.
In the presentation of our results we concentrate on polymers with $N$=100 
and 343, the former because the smaller size allows for clear visualization 
of the structures of the different phases, and the latter because 
the effects of finite size will be smallest. 

\begin{table}
\caption{Properties of the global optima for a 100-unit polymer.
$\epsilon_g^{min}$ and $\epsilon_g^{max}$ give the range of 
$\epsilon_g$ for which a structure is the global minimum.
$c/\protect\sqrt{ab}$ is the aspect ratio
and $\epsilon_g^{opt}$ is the value of the stiffness
for which a crystallite with that aspect ratio is expected to 
be lowest in energy and is given by 
$\epsilon_g^{opt}=
(c/\protect\sqrt{ab}-1)\epsilon/2$. }
\medskip
\begin{tabular}{ccccccc}
 & $n_{pp}$ & $n_g$ & $\epsilon_g^{min}/\epsilon$ & $\epsilon_g^{max}/\epsilon$ 
 & $c/\protect\sqrt{ab}$ & $\epsilon_g^{opt}/\epsilon$ \\
\noalign{\vskip 2pt}
\hline
\noalign{\vskip 2pt}
 A & 136 & 38 & 0.000 &  0.375 &  1.118 & 0.059 \\
 B & 133 & 30 & 0.375 &  0.500 &  1.563 & 0.281 \\
 C & 129 & 22 & 0.500 &  0.833 &  2.406 & 0.703 \\
 D & 124 & 16 & 0.833 &  2.167 &  3.704 & 1.352 \\
 E & 111 & 10 & 2.167 &  3.500 &  6.804 & 2.902 \\
 F &  97 &  6 & 3.500 & 12.000 & 12.500 & 5.750 \\
\end{tabular}
\label{table:100}
\end{table}

In order to monitor the orientational order within the polymer we devised 
an order parameter, $Q$, which is given by
\begin{equation}
Q={1\over N-1}\sqrt{{3\over 2}\sum_{\alpha=x,y,z} 
  \left(n_\alpha-{(N-1)\over 3}\right)^2},
\end{equation}
where $n_\alpha$ is the number of bonds in the direction $\alpha$ and $(N-1)/3$ 
is the expected value of $n_\alpha$ if the bonds are oriented isotropically.
$Q$ has a value of 1 if all the bonds are in the same direction, 
i.e.\ the polymer has a linear configuration, and a value of 0 
if the bonds are oriented isotropically. 

\begin{figure}
\epsfig{figure=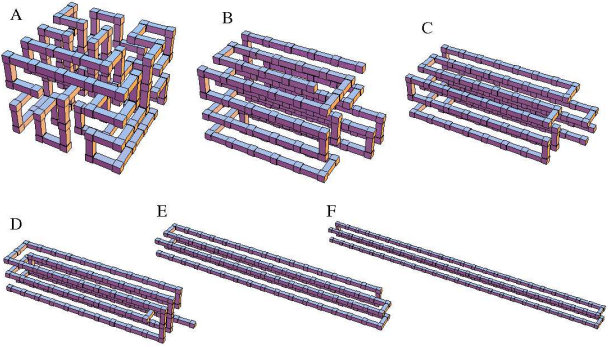,width=8.2cm}
%\vglue-0.7cm
\caption{Global minima of the 100-unit polymer at different values of 
$\epsilon_g$. The labels correspond to those in Table \ref{table:100}.}
\label{gmin}
\end{figure}

\section{Results}

\subsection{Solid phase}

In Table \ref{table:100} the properties of the global minima for $N$=100 are 
given for the range of $\epsilon_g$ we consider in this study, and examples of 
these minima are illustrated in Figure \ref{gmin}. 
The shapes of the global minima agree well with that expected based on 
the analysis of section \ref{sect:pmodel}.
In the final column of Table \ref{table:100} we have given the value of the 
stiffness for which a crystallite with the aspect ratio of the global minimum is 
expected to be lowest in energy based on Equation (\ref{eq:aspect}).
This value generally lies in the middle of the range for which the structure
is the global minimum.
Most of the global minima have a degeneracy associated with the different 
possible ways of folding the chain back and forth. 
This degeneracy decreases as the chains become stiffer 
and the crystallites become more extended. 
For $\epsilon_g=0$ there is no constraint on $n_g$, and so the global minimum has
a much larger  degeneracy, and the majority of the isomers of the global minimum 
have no orientational order. 

\begin{figure}
\epsfig{figure=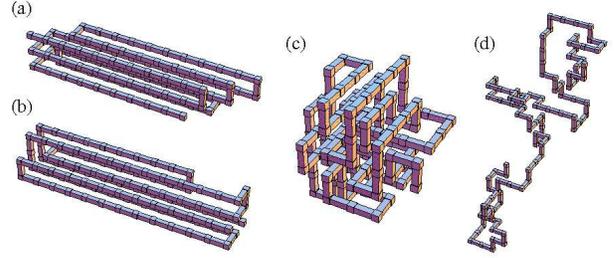,width=8.2cm}
%\vglue-0.3cm
\vspace{3mm}
\caption{Visualizations of various states of a 100-unit polymer. 
(a) A polymer based on structure D with some disorder in the stem lengths 
from a simulation at $T=0.9\epsilon k^{-1}$ and $\epsilon_g=2\epsilon$.
(b) A folded structure with 8 stems that contributes
to the middle peak of Figure \ref{100.ss}b 
($T=1.2\epsilon k^{-1}$, $\epsilon_g=3\epsilon$).
(c) A typical configuration of the dense globule with $R_g^2=5.60$
($T=0.75\epsilon k^{-1}$, $\epsilon_g=\epsilon$). 
(d) A typical configuration of the coil with $R_g^2=38.91$
($T=5.0\epsilon k^{-1}$, $\epsilon_g=\epsilon$) }
\label{pics}
\end{figure}

The global minimum is the free energy global minimum at zero temperature. 
However, as the temperature is increased it will become favourable to introduce
defects into the structures. 
One of the mechanisms we commonly observed was through fluctuations 
in the lengths of the folds, moving in and out like the slide of a trombone.
An example of such a structure is shown in Figure \ref{pics}a. 
The generation of this type of defect
is especially common at larger values of 
$\epsilon_g$ since it does not involve an increase of $n_g$.

\begin{figure}
\epsfig{figure=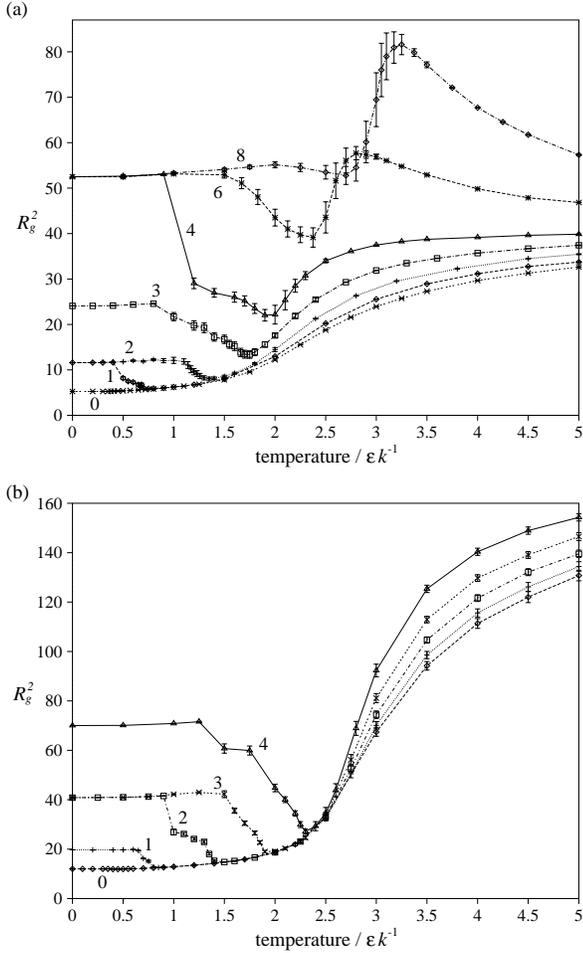,width=8.2cm}
\vspace{3mm}
\caption{Behaviour of $R_g^2$ as a function of temperature and $\epsilon_g$ for 
(a) $N$=100 and (b) $N$=343. Each line is labelled by the value of 
$\epsilon_g/\epsilon$.}
\label{rgsq}
\end{figure}

\begin{figure}
\epsfig{figure=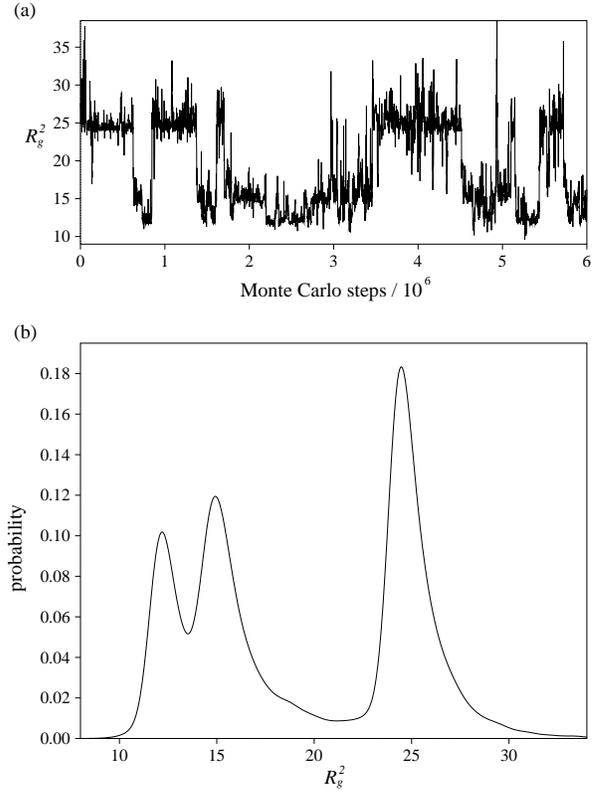,width=8.2cm}
\vspace{3mm}
\caption{Solid-solid coexistence observed for a 100-unit polymer with 
$\epsilon_g=3\epsilon$ at $T=1.2\epsilon k^{-1}$.
(a) Fluctuations in $R_g^2$ during a 10 million step Monte Carlo run. 
(b) Probability distribution of $R_g^2$.}
\label{100.ss}
\end{figure}

The dependency of the degeneracy, and therefore the entropy, 
on the aspect ratio of the crystallites raises the possibility of transitions
to crystallites with a smaller aspect ratio as the temperature is increased. 
Indeed, this is usually what is seen and leads to the decrease of $R_g^2$ with 
temperature that is observed before melting [Figure \ref{rgsq}].
At the centre of the transition, where the free energy of each crystallite is 
equal, ideally, the polymer would be seen to oscillate between the two forms 
[Figure \ref{100.ss}a] during the simulation spending equal time in each. 
This `dynamic coexistence' of structures leads to a bimodal 
(or even multimodal if more than two forms are stable) distribution of 
$R_g^2$ [Figure \ref{100.ss}b]. 
Since the Landau free energy is given by $A_L(q)=A-kT\log p_q(q)$, where $A$ is 
the Helmholtz free energy and $p_q(q)$ the canonical probability distribution 
for an order parameter $q$, the multimodality in $R_g^2$ implies that there
are free energy barriers between the different crystallites.
In the example shown in Figure \ref{100.ss}, the peak with the highest value 
of $R_g^2$ corresponds to structures similar to the global minimum D, 
and the peak with the lowest value of $R_g^2$ have structures similar to C. 
The other peak consists of structures that have eight aligned stems, either in
a $4\times 2$ array or a square array with one corner unoccupied 
[Figure \ref{pics}b].

However, isomerization between the different forms often requires a large-scale 
change in structure, which is particularly difficult if these transitions occur 
at low temperature, where the free energy barriers to transitions are largest.
With simulation techniques that only apply local moves, transitions between 
crystallites would be effectively impossible to observe
and even with a technique such as configurational-bias Monte Carlo, 
where global moves are possible, 
transitions may be rare, particularly for the larger polymers in this study.
This possible lack of ergodicity on the simulation time scales can lead 
occasionally to the abrupt jumps seen in $R_g^2$ 
(e.g.\ for $N$=100 at $\epsilon_g$=4$\epsilon$ and $T$=1$\epsilon k^{-1}$,
and for $N$=343 at $\epsilon_g$=2$\epsilon$ and $T$=0.95$\epsilon k^{-1}$) 
rather than the smoother transition that would 
be expected if equilibrium values of $R_g^2$ were obtained.

The differences in energy and entropy between crystallites of
different aspect ratio are due to surface effects and so scale less than 
linearly with size. Therefore, these solid-solid transitions are not finite-size
analogues of {\it bulk} first-order phase transitions.

Interestingly, this coexistence of polymers with different cuboidal shapes but 
the same basic structure bears some resemblance to the coexistence of cuboidal 
sodium chloride clusters that has recently been observed 
experimentally.\cite{Hudgins97a} All the clusters have the rock-salt structure 
but the cuboids have different dimensions.

\begin{figure}
\epsfig{figure=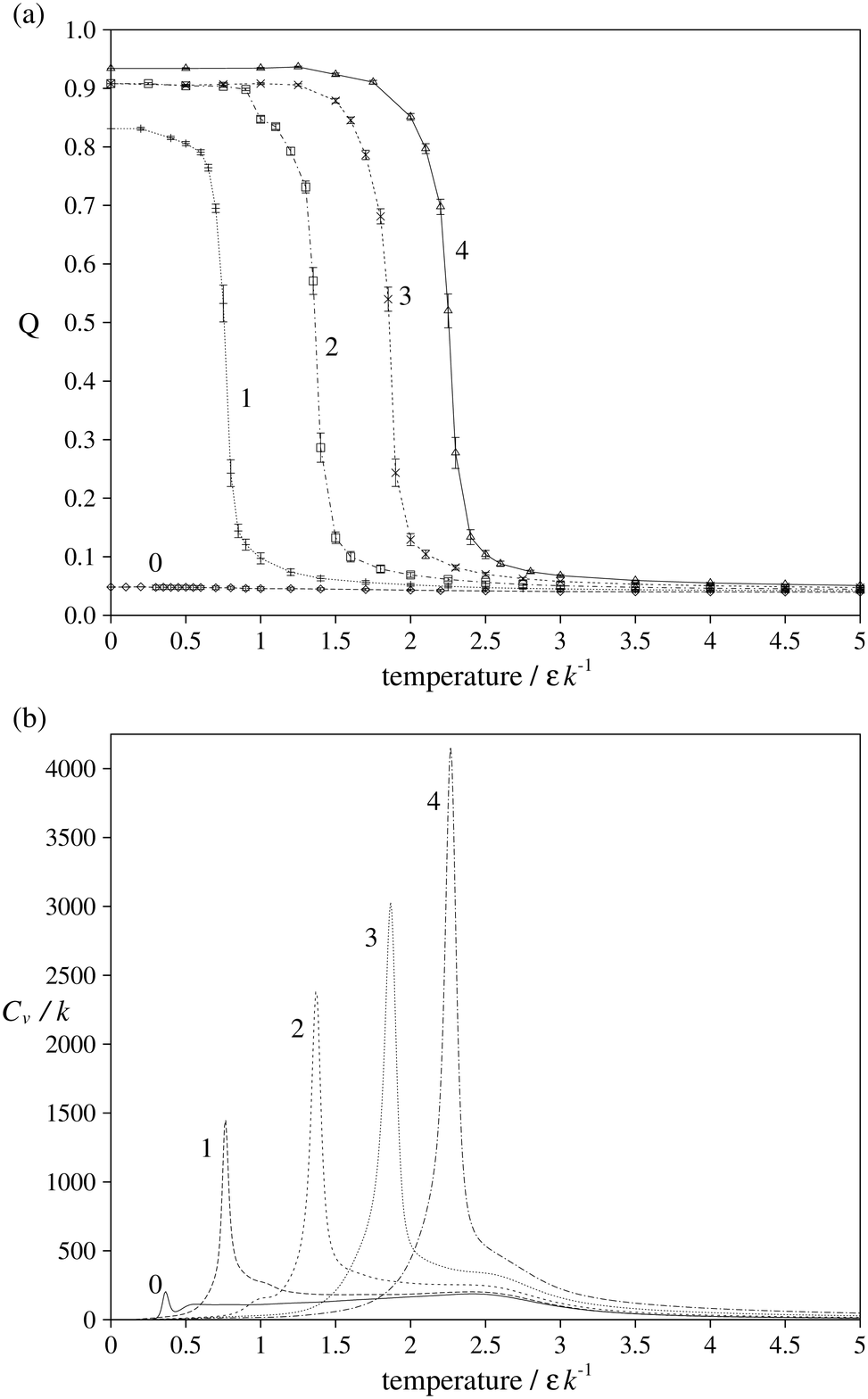,width=8.2cm}
\vspace{3mm}
\caption{Behaviour of (a) $Q$ and (b) $C_v$ as a function of temperature and 
$\epsilon_g$ for $N$=343. 
Each line is labelled by the value of $\epsilon_g/\epsilon$.}
\label{opcv}
\end{figure}

\subsection{Solid-globule/coil transition}
\label{sect:sg}
For $\epsilon_g>0$ the global minimum is orientationally ordered with 
a value of $Q$ near to unity. 
Of course, this folded state corresponds to the free energy global minimum
at zero temperature.
However, as the temperature is increased there must come a point
when a disordered higher entropy state, be it the globule or coil,
becomes lowest in free energy, and the polymer loses its orientational
order---it `melts'.
This transition is signalled by a decrease in $Q$ to a value close to zero,
giving rise to a typical sigmoidal shape for the temperature dependence of $Q$ 
[Figure \ref{opcv}a]. This transition is accompanied by a peak in the heat 
capacity [Figure \ref{opcv}b], and we use the position of this maximum to 
define the melting temperature of the polymer, $T_m$. 
(Alternative definitions, such as the temperature at which 
$Q=0.5$, give practically identical results).
The transition also often involves a change in the radius of gyration, the sign 
and magnitude of this change depending on the stiffness of the polymer. 
For lower values of $\epsilon_g$ the transition to the dense {\it globule}, 
leads to a decrease in $R_g$, but at higher values of $\epsilon_g$ the 
transition to the {\it coil\/} leads to an increase in $R_g$
($\epsilon_g$=6$\epsilon$, 8$\epsilon$ in Figure \ref{rgsq}a).

\begin{figure}
\epsfig{figure=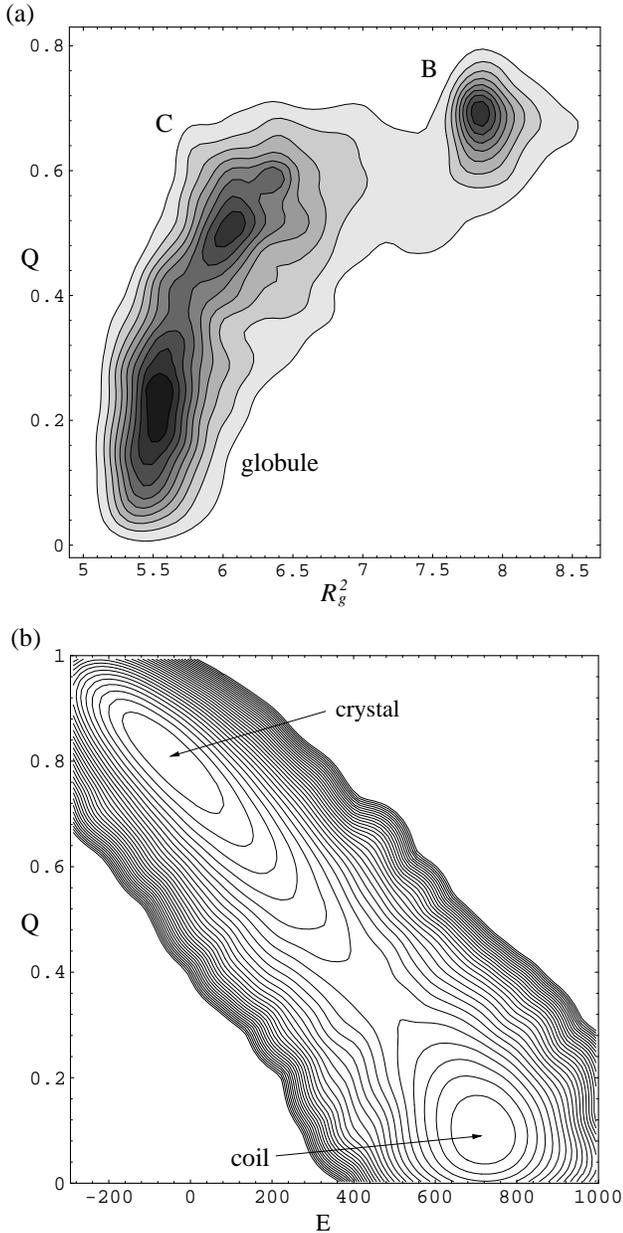,width=8.2cm}
\vspace{3mm}
\caption{(a) Solid-globule coexistence observed for a 100-unit polymer with 
$\epsilon_g=\epsilon$ at $T=0.675\epsilon k^{-1}$. 
Two-dimensional probability distribution in $Q$ and $R_g^2$.
The labels $B$ and $C$ refer to the global minima of Table \ref{table:100} on
which the structure of the polymers contributing to the maxima are based.
(b) Crystal-coil coexistence for a 343-unit polymer with
$\epsilon_g=\epsilon$ at $T=3.45\epsilon k^{-1}$.
Two-dimensional free energy profile in $Q$ and $E$. 
The contours occur at intervals of $0.5kT$ above the global free energy minimum. 
The contours over $15kT$ above the global minimum are not plotted.}
\label{100.sl}
\end{figure}

At temperatures in the transition region, as for the solid-solid transitions,
dynamic coexistence of the ordered and disordered states is seen, leading to 
multimodal probability distributions (and free energy barriers) 
[Figure \ref{100.sl}].
In the example shown in Figure \ref{100.sl}a three states are seen. 
Structures based on the global minima B and C give rise to the two ordered states. 
The low value of $Q$ ($\sim 0.5$) for the state associated with structure $C$ is 
a reflection of the considerable disorder that can be present in the solid 
state at $T_m$. 
The peak with the lowest $Q$ value is due to the disordered globule and 
an example of a polymer configuration that contributes to this peak is given in 
Figure \ref{pics}c.

\begin{figure}
\epsfig{figure=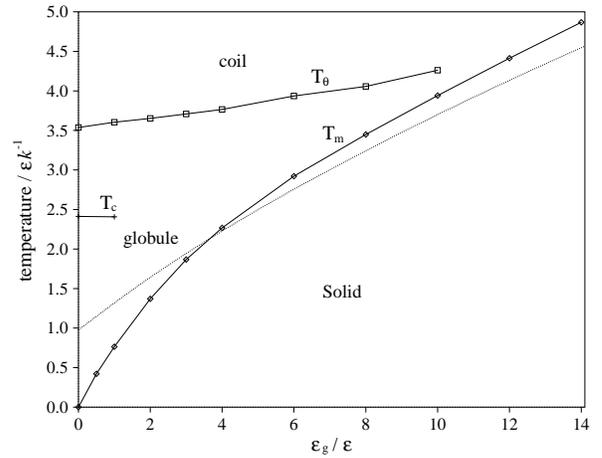,width=8.2cm}
\vspace{3mm}
\caption{Phase diagram of the 343-unit polymer. 
The phase diagram is divided into regions by the values of $T_m$ and $T_\theta$. 
The position of the heat capacity peak associated with collapse, $T_c$,
has also been included.
The solid lines with data points are simulation results, and 
the dotted line is from the simple theoretical calculation of $T_m$
outlined in Section \ref{sect:sg}.}
\label{phase}
\end{figure}

The effects of stiffness on this order-disorder transition can be seen from 
Figures \ref{opcv} and \ref{phase}.
The temperature of the transition increases with stiffness, and
at larger values of $\epsilon_g$ this increase is a little slower than linear.
Accompanying this change is an increase in the height of the heat capacity peak, 
and therefore the latent heat of the transition.
This increasing energy gap between the ordered and disordered states is a 
result of the increasing energetic penalty for the larger number of gauche 
bonds associated with the disordered state.
As $T_m=\Delta E/\Delta S$, the dependence of the energy gap, $\Delta E$, on the 
stiffness is one of the main causes of the increase in $T_m$. 
This effect is reinforced by the changes in the entropy difference between 
ordered and disordered states, $\Delta S$:
as $\epsilon_g$ increases the number of gauche bonds decreases, thus lowering 
the number of configurations contributing to the disordered state.

This behaviour is similar to simple liquids where decreasing the range of 
the potential leads to an increasing energy gap between solid and liquid because 
of the increasing energetic penalty for the disorder associated with the 
dispersion of nearest-neighbour distances in the liquid, thus playing an 
important role in the destabilization of the liquid phase.\cite{Doye96a,Doye96b}

The respective roles of energy and entropy in the increase of $T_m$ with 
stiffness can be investigated more quantitatively by approximating the free 
energy of the disordered state by that for the ideal coil. 
This is a reasonable approximation, since 
the ideal expression for $n_g$, $n_g^{ideal}=4N/(\exp(\beta\epsilon_g)+4)$,
fits the simulation values for the coil, and for the globule, fairly closely.
The free energy of an ideal coil $A_{ideal}$ is
\begin{equation}
A_{ideal}=-N k T\log\left( 1+ 4 \exp(-\beta \epsilon_g)\right).
\label{aid}
\end{equation}
Decomposing this expression into its energetic and entropic components
showed that the increase in the energy with stiffness is 
the main contributor to the change in the free energy of the disordered 
state with stiffness, except at the higher temperatures 
relevant to the `melting' transition for larger values of $\epsilon_g$.

Use of Equation (\ref{aid}) also allows us to calculate a value of $T_m$ if
we approximate the free energy of the solid by only its energetic component 
[Equation (\ref{eq:lbound})].
This simple calculation gives surprisingly good agreement with the simulation 
data [Figure \ref{phase}], especially at larger $\epsilon_g$. 
It breaks down at low $\epsilon_g$, e.g.\ the prediction of a non-zero value 
of $T_m$ at $\epsilon_g$=0, because Equation (\ref{aid}) is the free energy 
of a coil in the absence of any interactions; 
for the dense polymers at low $\epsilon_g$ and near to $T_m$, the polymer-polymer 
contacts make a significant contribution to the energy of the disordered state.
The number of polymer-polymer contacts in the disordered polymer decreases 
significantly with increasing temperature, because of the larger entropy of 
less dense configurations. 
Therefore, Equation (\ref{aid}) becomes a better approximation to the free energy 
at higher temperatures, and thus provides a better description
of melting at the higher temperatures relevant for larger $\epsilon_g$.

Our simulation results for $T_m$ are also in qualitative agreement with 
theoretical results in which a more sophisticated treatment of the 
globule\cite{Doniach96a} gives the correct behaviour for $T_m$ at 
low $\epsilon_g$. 
However, a drawback of the description of Doniach {\it et al.\/} is that the 
free energy of the coil per polymer unit is nonzero in the limit of large 
$\epsilon_g/kT$, causing $T_m$ to reach an asymptotic value, 
rather than continuing to increase with $\epsilon_g$.

The increase in $\Delta E$ with stiffness also has an effect on the coexistence 
of ordered and disordered states.
At large values of $\epsilon_g$ there is multimodality in the 
probability distribution for the energy as well as for the order parameter, 
because of larger free energy barriers between the states. 
In the example shown in Figure \ref{100.sl}b there is a free energy barrier of 
$3.24 kT$ for passing from the crystal to the coil.

\begin{figure}
\epsfig{figure=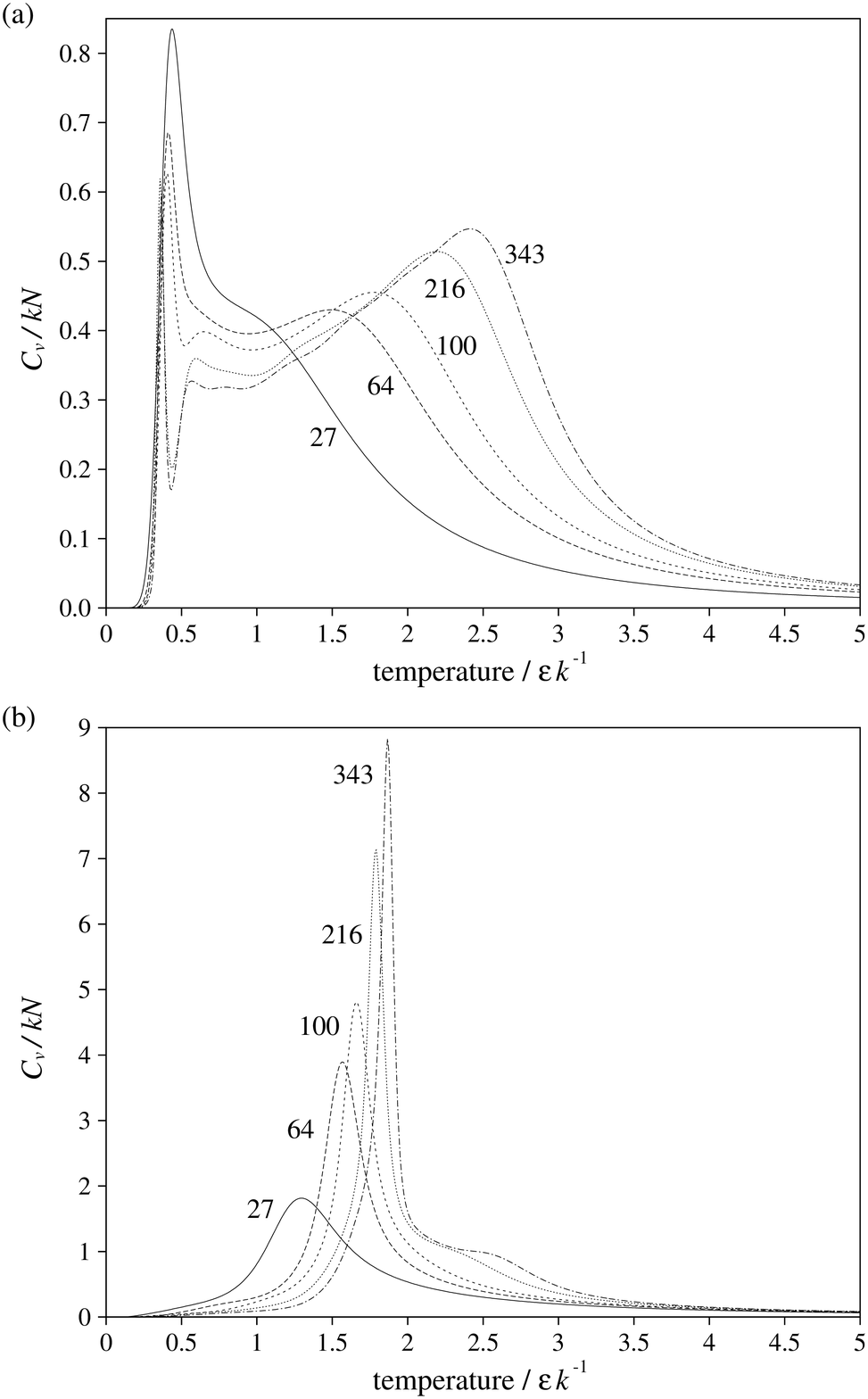,width=8.2cm}
\vspace{3mm}
\caption{$C_v/N$ as a function of size for (a) $\epsilon_g=0$ and 
(b) $\epsilon_g=3\epsilon$.}
\label{cv.scale}
\end{figure}

A consideration of the effect of size on the transition
shows that as the polymer becomes longer the melting point becomes higher, 
and the transition becomes sharper with an increasing latent heat per monomer 
[Figure \ref{cv.scale}b].
This is consistent with the transition being the finite-size analogue of a 
first-order phase transition. $T_m$ increases with size because the effect of 
the surface term (the $N^{2/3}$ term) in the
energy of a crystallite [Equation (\ref{eq:lbound})] diminishes with size.
Moreover, as the coefficient of the surface term increases with $\epsilon_g$ 
(the higher aspect ratio crystallites have a larger surface area)
the effects of size on $T_m$ are more pronounced at larger $\epsilon_g$.

At $\epsilon_g$=0 the behaviour is qualitatively different because there is 
no energy difference between the orientationally ordered and disordered forms. 
As there are far fewer states that possess orientational order, they are never 
thermodynamically favoured and so there is no orientational order-disorder 
transition and $Q$ always has a low value [Figure \ref{opcv}a].
Despite this, the polymers we studied do have a low temperature heat capacity
peak at $T\sim 0.4\epsilon k^{-1}$ [Figure \ref{cv.scale}a].
This feature stems from a transition between the maximally compact cuboidal 
global minimum and a more spherical dense globule, and gives rise to bimodality 
in the canonical probability distribution of the energy. 
However, the latent heat per atom for the transition decreases with increasing 
size, indicating that the transition is not tending to a first-order phase 
transition in the bulk limit. 
Furthermore, the form of the heat capacity can be very different for sizes
at which it is not possible to form complete cuboids.

The order-disorder transition observed by Zhou {\it et al.\/} was 
also for a fully flexible polymer.\cite{Zhou96a} 
However, as the model used was off-lattice, unlike in our model, there
is the possibility of a first-order transition associated with the 
condensation of the monomers onto a lattice.

\subsection{The globule to coil transition}
\begin{figure}
\epsfig{figure=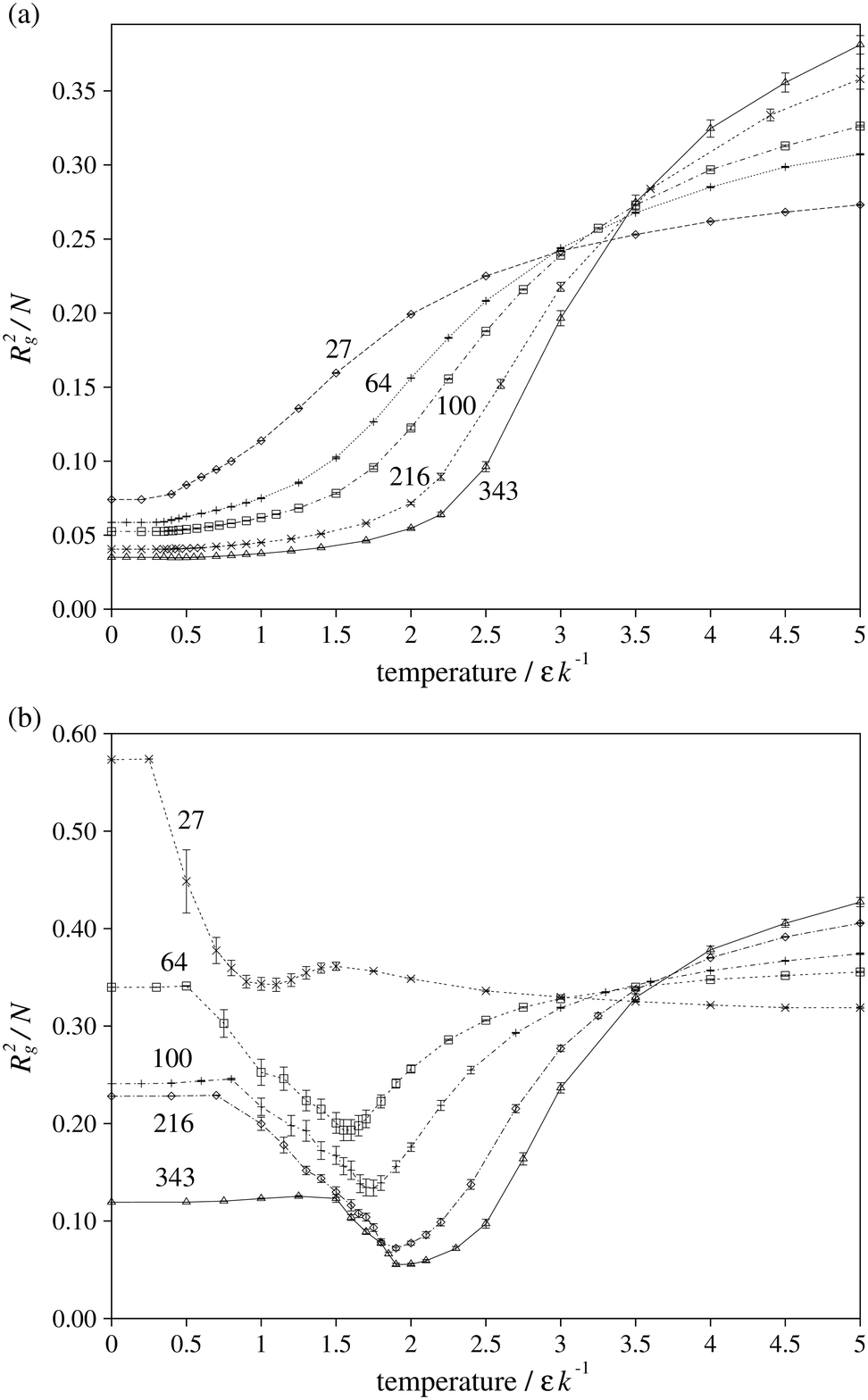,width=8.2cm}
\vspace{3mm}
\caption{$R_g^2/N$ as a function of size for (a) $\epsilon_g=0$ and 
(b) $\epsilon_g=3\epsilon$.}
\label{rg.scale}
\end{figure}

For polymers, there can be two disordered phases, the dense globule 
[Figure \ref{pics}c] and the coil [Figure \ref{pics}d]. They are 
differentiated by the scaling behaviour of the size of the polymer in each phase.
For the globule $R_g^2\propto N^{2/3}$ and for the coil $R_g^2\propto N^{6/5}$. 
Between these two states is the $\theta$-point where $R_g^2\propto N$ and 
the polymer is said to behave ideally.
On passing from the coil to the globule the polymer collapses,
as seen in the sigmoidal shapes of the $R_g^2$ curves [Figure \ref{rg.scale}a].
The collapse transition can also give rise to a high temperature peak in the 
heat capacity, which is more rounded than that for melting and the latent 
heat of which is associated with the loss in polymer-polymer contacts on going 
to the lower density coil [Figure \ref{cv.scale}a].
These features are most clear at $\epsilon_g=0$ where they are not obscured by
the features due to the melting transition.
At higher values of $\epsilon_g$ the sigmoidal shape of $R_g^2$ is cut off at
low temperatures by the rise 
associated with the transition to the crystalline states [Figure \ref{rgsq}],
and the collapse transition just causes a high temperature shoulder 
in the heat capacity [Figure \ref{opcv}b]. 
As the size of the polymer increases the transition becomes sharper 
with a steeper rise in $R_g^2$ [Figure \ref{rg.scale}a] and 
a narrower heat capacity peak at a higher temperature [Figure \ref{cv.scale}a]. 
In the limit $N\rightarrow\infty$ this heat capacity peak 
occurs at the $\theta$-point.

To complete the phase diagram for our model polymer we estimated 
$T_\theta$ by making use of the scaling relations: 
a series of plots of $R_g^2/N$ should all cross at $T_\theta$. 
However, due to finite size effects the value obtained by this method only 
approaches the exact $T_\theta$ (from below) as $N\rightarrow\infty$. 
For example, we estimate $T_\theta(\epsilon_g$=0)=3.537 from the crossing 
point of the lines for $N$=216 and 343,
whereas an accurate determination gives $T_\theta$=3.721.\cite{Grassberger95a}

Figure \ref{phase} shows that $T_\theta$ has only a weak dependence on 
$\epsilon_g$. 
Combined with the increase of $T_m$ with $\epsilon_g$, 
this leads to the loss of the globular phase at $\epsilon_g\sim 12\epsilon$. 
Above this value the polymer passes directly between the solid and 
the coil phases, i.e.~there is only one disordered phase.
An analogy can be made to the phase behaviour of simple atomic systems, 
with a correspondence between the dense globule and the liquid and 
between the coil and the vapour.
For both systems the denser phase disappears as an interaction parameter 
is varied, for polymers as the stiffness is increased and 
for atomics systems as the range of the potential is decreased.\cite{Hagen94a} 

Our phase diagram is consistent with the work of Bastolla and Grassberger. 
From simulations of polymers longer than those we study here, they found that 
the globular phase disappears at $\epsilon_g\sim 13\epsilon$.\cite{Bastolla97a}
The phase diagrams is also very similar to the theoretical predictions of
Doniach {\it et al.\/} who estimate the loss of the globular phase to occur 
at $\epsilon_g\sim 15\epsilon$.\cite{Doniach96a} 
However, they assumed that $T_\theta$ is constant whereas both in 
ref.~\onlinecite{Bastolla97a} and here a small increase with 
$\epsilon_g$ is observed.

Precursors of the loss of the globular phase can be seen at $\epsilon_g<12$. 
These effects are due to finite size, and for smaller polymers they
occur at lower values of $\epsilon_g$. 
Firstly, the features in the heat capacity due to collapse become engulfed
by the peak due to melting. As the stiffness increases,
the collapse peak becomes only a shoulder and then 
it disappears altogether [Figure \ref{opcv}b]. 
Secondly, on melting the polymer can pass directly from the solid to 
a low density disordered state, albeit one where $R_g^2$ scales 
less than linearly with size. 
This effect can be seen for the 27-unit polymer at 
$\epsilon_g=3\epsilon$ [Figure \ref{rg.scale}], and is also seen for 
longer polymers at larger values of $\epsilon_g$ [Figure \ref{rgsq}a].
It results from the expansion of sufficiently stiff and small polymers 
with decreasing temperature, which is because the persistence length 
becomes a significant fraction of the total length.
An estimate of the persistence length is given by the ratio $N/n_g$, the
average length between gauche bonds, $N/n^{ideal}_g=1+\exp(\beta\epsilon_g)/4$.

Considerable theoretical effort has gone into studying the evolution
of the coil--globule transition as the stiffness of the polymer is increased,
see Refs.\ \onlinecite{Grosberg92a,Williams81a} and references therein.
The theories predict that the coil--globule transition is second order for 
flexible polymers (this has been confirmed by simulation\cite{Grassberger95a})
but becomes first order as the stiffness increases. Due to the
small size of the polymers we have simulated 
we are unable to determine the order of the transition accurately and
so cannot test this prediction.
However, the theories all assume a liquid-like dense phase; they neglect the
possibility of an ordered dense phase, despite
it being well known that the dense state of DNA, a stiff polymer, has
hexagonal order.\cite{Lerman71a,Ubbink95a}
It is clear from Figure \ref{phase} that for stiff polymers, the coil--globule 
transition is preempted by a coil--solid transition.

The difference in phase behaviour between flexible and stiff polymers can
be easily understood.
As a flexible polymer is cooled, the Boltzmann weight for polymer-polymer 
contacts increases and so the number of contacts increases, at a cost in the 
entropy of the polymer. The entropy cost derives from the fact that the polymer
must bend back on itself in order for the units to be in contact. The increase
in the number of contacts is continuous and so the radius of gyration of the
polymer varies continuously --- the coil--globule transition is second order.
However, bending a stiff polymer back on itself is more difficult, it costs both
entropy and energy. This larger cost can be repaid if more than one pair of units
is in contact, which is true if the two parts of the polymer run parallel to each
other for several units. Of course, as the polymer is stiff the entropy cost
for two parts of the polymer to run parallel for a number of units is small.
The low energy configurations of a stiff polymer are those with long parts of 
the polymer parallel, see Section IIA. The energy gain is even larger if a 
number of parts of the polymer form a bundle,\cite{Sear97a,Sear97b}
e.g. if four parts of the polymer which are running parallel form a square 
bundle the energy is not twice but four times that of two parts running parallel.
So, when the polymer is cooled below the point where the energy gain of bundles
outweighs their entropy cost then these bundles proliferate and the 
radius of gyration drops suddenly --- the coil--solid transition is first order.

Our results can also be related to experiment.
The coil-globule transition of polystyrene, a flexible polymer, is continuous
and the globule is liquid-like.\cite{Grosberg92a,Sun80a}
For DNA, an example of a stiff polymer,
a first-order transition has been observed between the coil and a
compact dense state,\cite{Lerman71a,Ueda96a}
which has hexagonal order.\cite{Lerman71a,Ubbink95a}
The lattice model used here also has a continuous \cite{Grassberger95a}
coil--globule transition when $\epsilon_g/\epsilon$ is small
and a first-order transition from a coil
to an ordered dense state when $\epsilon_g/\epsilon$ is large.
It is thus able to reproduce the phenomenology of the existing experimental data.
However, we do not know of any examples where the three phases---coil, globule 
and crystalline---predicted by our model to occur for semi-flexible polymers 
have been experimentally observed.

\section{Conclusion}

In this paper we have provided an example of an order-disorder transition that 
can be observed for an isolated homopolymer. Its basis in the interactions of 
the model is clear.
The global minimum is orientationally ordered in order to minimize the 
energetic penalty for gauche bonds. 
At some finite temperature this order must be lost in a transition to the 
higher entropy globule or coil. 
As the stiffness of the polymer increases the energy gap between the
ordered and disordered states increases contributing to an increase in 
the temperature at which this `melting' transition occurs.
This effect, coupled with the weak dependence of the $\theta$-point on 
the stiffness parameter, leads to the disappearance of the disordered globule 
for sufficiently stiff polymers; there is 
no longer a collapse transition from the coil to globule. 
This behaviour is analogous to the disappearance of the liquid phase of simple 
atomic systems as the range of the potential is decreased. 
The phase diagram we obtain is in good agreement with recent 
theoretical\cite{Doniach96a} and simulation\cite{Bastolla97a} 
studies of this polymer model.

One of the intriguing aspects of our model is the folded structures 
that form at low temperature. 
Firstly, these states are not artifacts of our lattice model. 
In simulations of isolated polyethylene chains,
for the same energetic reasons as in our model---the extra energy from 
polymer-polymer contacts outweighs the energetic cost of forming a 
fold---relaxation to folded structures was 
observed.\cite{Kavassalis93a,Sundararajan95a}
Furthermore, as here, it was also found that the aspect ratio of the 
crystallites increased with increasing stiffness.\cite{Sundararajan95a}
However, we know of no experiments on a homopolymer system where evidence has 
been found for a transition from a disordered globule to a crystalline 
structure with chain folds. 
The most likely signature for such a transition would be an increase in the 
radius of gyration with decreasing temperature in very dilute polymer solutions.

It is also natural to ask whether this study can provide any insights into 
polymer crystallization, since polymer crystals have a lamellar morphology 
in which the polymer is folded back and forth in a similar manner to 
the folded structures we observe. 
However, the formation of the folded structures for the isolated polymer is a 
purely thermodynamic effect---the structures are the global minima---whereas in 
the bulk case it is a kinetic effect\cite{Hoffman76a,Armistead92a}---it
is generally accepted that the global minimum is a crystal where all the chains
are in an extended conformation.
The study, though, may have a more indirect relevance to the crystallization 
of polymers from solution.
Although little consideration has been given to the structure of the
polymer arriving at the surface in theories of polymer crystallization, 
it is not implausible that the existence of significant ordering in
the adsorbing polymer would have a considerable effect on the 
crystallization process.

Although the present study only examines the homopolymer thermodynamics,
some comments concerning the dynamics can be made. 
Klimov and Thirumalai made the interesting observation that
model proteins are most likely to be good folders when 
$T_f/T_\theta$ is large, where $T_f$ is the temperature at 
which the transition to the native state of the protein occurs.\cite{Klimov96a}
For our system $T_m/T_\theta$ increases with $\epsilon_g$
[Figure \ref{phase}]. Therefore, if Klimov and Thirumalai's relation 
also holds for our homopolymers, one would expect crystallization of 
the polymer to become more rapid as the stiffness increases---crystallization 
is easier direct from the coil than via the disordered dense globule.
However, whereas at $T_f$ there is a transition to a single state, 
the globule minimum,
at our $T_m$ a transition to an ensemble of crystalline structures occurs.
Therefore, it does not necessarily follow that the homopolymers would be 
able to reach the global minimum more rapidly as the stiffness increased---indeed
it might be that trapping in low-lying crystalline structures is more 
pronounced for stiff polymers.

\acknowledgements
The work of the FOM Institute is part of the research program of
`Stichting Fundamenteel Onderzoek der Materie' (FOM) and is 
supported by NWO (`Nederlandse Organisatie voor Wetenschappelijk Onderzoek'). 
JPKD acknowledges the financial support provided by the Computational 
Materials Science program of the NWO, 
and RPS would like to thank The Royal Society for the award of a fellowship.
We also thank Mark Miller for use of a program to perform the multi-histogram 
analysis.

\end{document}